\documentclass{epsconf}
\usepackage{graphicx}
\usepackage{epsfig} 
\usepackage{wrapfig}
\usepackage{amsmath}
\usepackage{siunitx}

\usepackage{authblk}

\newcommand{\etal}{\textit{et al.}}
\title{Enhanced betatron-radiation energy using two collinear laser pulses }

\author[1]{\underline{Z. M. Chitgar}}
\author[1,2]{P. Gibbon}
\author[3]{J. B\"oker}
\author[3,4]{A. Lehrach}
\author[5,6]{M. B\"uscher}

\affil[1]{Institute for Advanced
	Simulation, J\"ulich Supercomputing Centre, Forschungszentrum J\"ulich, D-52425 J\"ulich,	Germany}
\affil[2]{Centre for Mathematical Plasma Astrophysics, KU Leuven, 3000 Leuven, Belgium}
\affil[3]{Institut f\"ur Kernphysik (IKP-4), Forschungszentrum J\"ulich, D-52425 J\"ulich, Germany}
\affil[4]{III. Physikalisches Institut B, RWTH Aachen, Germany}
\affil[5]{Peter Gr\"unberg Institut (PGI-6), Forschungszentrum J\"ulich GmbH,  D-52425 J\"ulich, Germany}
\affil[6]{Institut f\"ur Laser- und Plasmaphysik HHU D\"usseldorf, D-40225 D\"usseldorf, Germany}

\begin{document}
\maketitle

\vspace{-1.5cm}A new self-injection scheme is proposed for the laser wakefield accelerator in the nonlinear (bubble) regime using a pair of matched, copropagating laser pulses which yields a low emittance, \SI{}{\pico\coulomb} electron bunch. By tuning their relative delay and intensity, the subsequent betatron radiation energy can be considerably ($\times$3) enhanced compared to the single pulse scheme for the same total energy. A general condition for the optimal bubble size is derived and verified by particle-in-cell simulations, further demonstrating the advantages of the double-pulse scheme for self-injection. Previously, multi-pulse schemes have been used as a means to achieve ionization injection of electrons in higher energy regimes~\cite{FebMulti}.

The bubble regime of electron acceleration~\cite{1ref} is the highly non-linear regime of laser wake field acceleration where the laser pulse intensity is high enough to create an electron-free cavity. Some electrons get trapped in the cavity, are accelerated and start to wiggle around the laser pulse propagation axis, resulting in betatron radiation~\cite{rousse}.  In the wiggler regime, the number of emitted photons and photon energy respectively, per electron and period are $N_{\gamma}=4.40\times 10^{-12} \sqrt{\gamma n_{e}[\SI{}{\per\cubic\centi\meter}]}r_{\beta}[\SI{}{\micro\meter}]$ and $\hbar\omega_c [\SI{}{\electronvolt}]=5.24 \times 10^{-21} \gamma^2 n_e [\SI{}{\per\cubic\centi\meter}] r_{\beta}[\SI{}{\micro\meter}]$~\cite{corde}; where $N_{\gamma}$ and $\hbar\omega_c $ are the number and maximum energy of the emitted X-ray photons, respectively; $n_e$ is the number density of the accelerated electrons and $r_{\beta}$ is the wiggle amplitude.


With two consecutive laser pulses, higher energy gain for electrons, higher beam current~\cite{horny} and enhanced betatron emission in the nonlinear blowout regime is possible. The double-cavity scheme achieves this firstly by ensuring that the plasma is fully ionized by the leading pulse; secondly, the accumulation and recycling of free electrons at the back of the first cavity provides a concentrated source for the second pulse to enhance the accelerating field behind it. The two separate pulses each carry a fraction of the total energy which would normally be contained in a single pulse. To make more quantitative predictions for the double-pulse scheme we have performed 2D particle-in-cell simulations using the EPOCH code~\cite{arber}, with the aim of finding the optimum relative delay between two pulses and the most effective relative energy fraction. Then, the X-ray energy spectrum is compared between the double- and single-pulse schemes.

All simulations were performed using a $100\times \SI{40}{\micro\meter\squared}$ box filled with helium gas of density $\num{9.2E17}\SI{}{\per\cubic\centi\metre}$, discretised by a computational grid with dimensions $n_x\times n_y = 3000\times 1200$. The target starts with a vacuum region of \SI{5}{\micro\meter} followed by a \SI{7}{\micro\meter} ramp in gas density in order to avoid a too steep gradient at the plasma edge. A $\SI{2}{\joule}$, $\SI{20}{\femto\second}$ laser with wavelength  $\SI{800}{\nano\metre}$ was focused from the left hand boundary down to $\SI{10}{\micro\metre}$ at the box centre, giving a nominal (single-pulse) intensity of $\num{3.184E19}\SI{}{\watt\per\centi\metre\squared}$, or $a_0 = 3.85$. A moving window was deployed in order to follow the development of the ensuing plasma wake and electron trapping.

\begin{wrapfigure}{}{70mm}\centering
	\vspace{0.0cm} 
	\includegraphics[width=70mm]{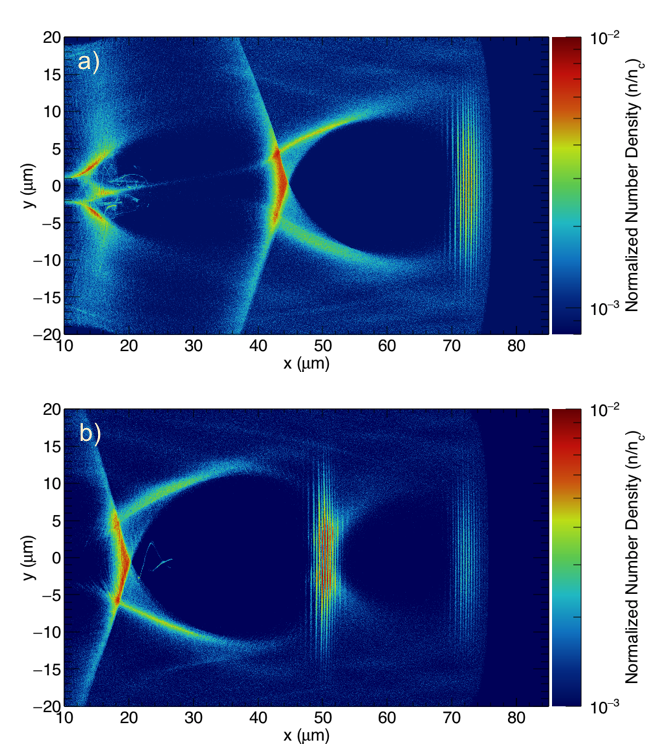}
	\caption{\it \small 2D snapshot of the electron number density distribution at t=\SI{1500}{\femto\second}, of the helium target being irradiated by a laser of $a_0 = 3.85$, for a) the single pulse scheme, b) the double pulse scheme with the optimum condition. The total pulse energy is \SI{2}{\joule}.}
	\label{numdens}
	\vspace{0cm} 
\end{wrapfigure}

\begin{wrapfigure}{}{70mm}\centering
	\vspace{0.0cm} 
	\includegraphics[width=70mm]{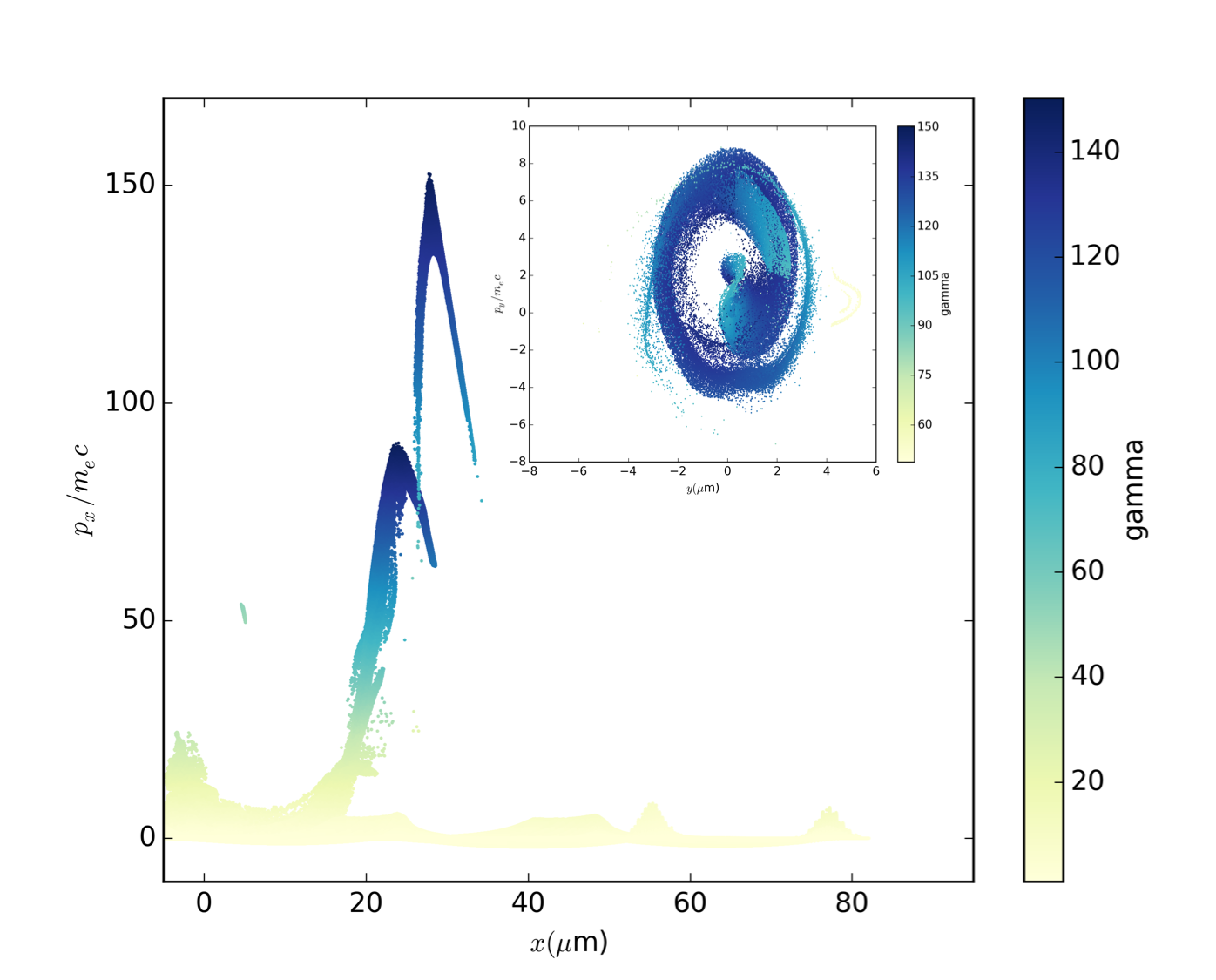}
	\caption{\it \small Comparison of electron momentum phase spaces $(x, p_x)$ at $t=\SI{1500}{\femto\second}$. The original pulse energy of the laser (2J) is divided between pulses at a ratio of 1:2, the second pulse carrying more, and a relative delay of 78 fs for the simulation with double pulses. The lower left bunch belongs to single-pulse and the upper right to the double-pulse scheme. Inset: beam emittance in the double scheme with $\epsilon_{rms}=\SI{11.9\pi}{\milli\metre\milli\radian}$.}
	\label{scat}
	\vspace{0cm} 
\end{wrapfigure}

A representative example of the new scheme is depicted in Fig.~\ref{numdens}, which compares the electron number density after $\textrm{t}=\SI{1500}{\femto\second}$ for the single and double-pulse schemes respectively, corresponding to approximately one Rayleigh length ($z_R=\pi w_0^2/\lambda$) of laser propagation. In the double-pulse scheme it is immediately apparent that the second cavity is larger and that consequently, a longer acceleration length is obtained. According to Fig.~\ref{numdens}-b), for the case of double pulses, a clean bunch of oscillating electrons is injected into the second cavity. With the same laser parameters and target characteristics, very few electrons are trapped in the first bubble in the single pulse scheme (Fig.~\ref{numdens}-a), a result consistent with the self-injection threshold predicted by the theoretical model of Ref.~\cite{Benedetti}. Therefore, in the remainder of this paper the trapped electrons in the wake behind the primary bubble are considered alongside the injected ones in double scheme in order to compare their energy distributions and momentum phase space. Figure~\ref{scat} depicts the electron distribution in momentum phase space $(x,p_x)$, confirming that no electrons are trapped in the first bubble in both schemes for these laser parameters. A significant share of the energy spectrum in the standard single-pulse scheme is taken up by electrons which are trapped in the wake behind the primary bubble (the left bunch). However, even though the aggregate laser energy is the same, the trapped electrons in the second cavity of the double-pulse scheme carry nearly twice that in the single-pulse case, with $\sim \SI{5}{\pico\coulomb}$ charge and emittance of $\epsilon_{rms}=\SI{11.9\pi}{\milli\metre\milli\radian}$ (Fig.~\ref{scat} inset).

\begin{wrapfigure}{}{70mm}\centering
 	\vspace{0.0cm} 
 	\includegraphics[width=70mm]{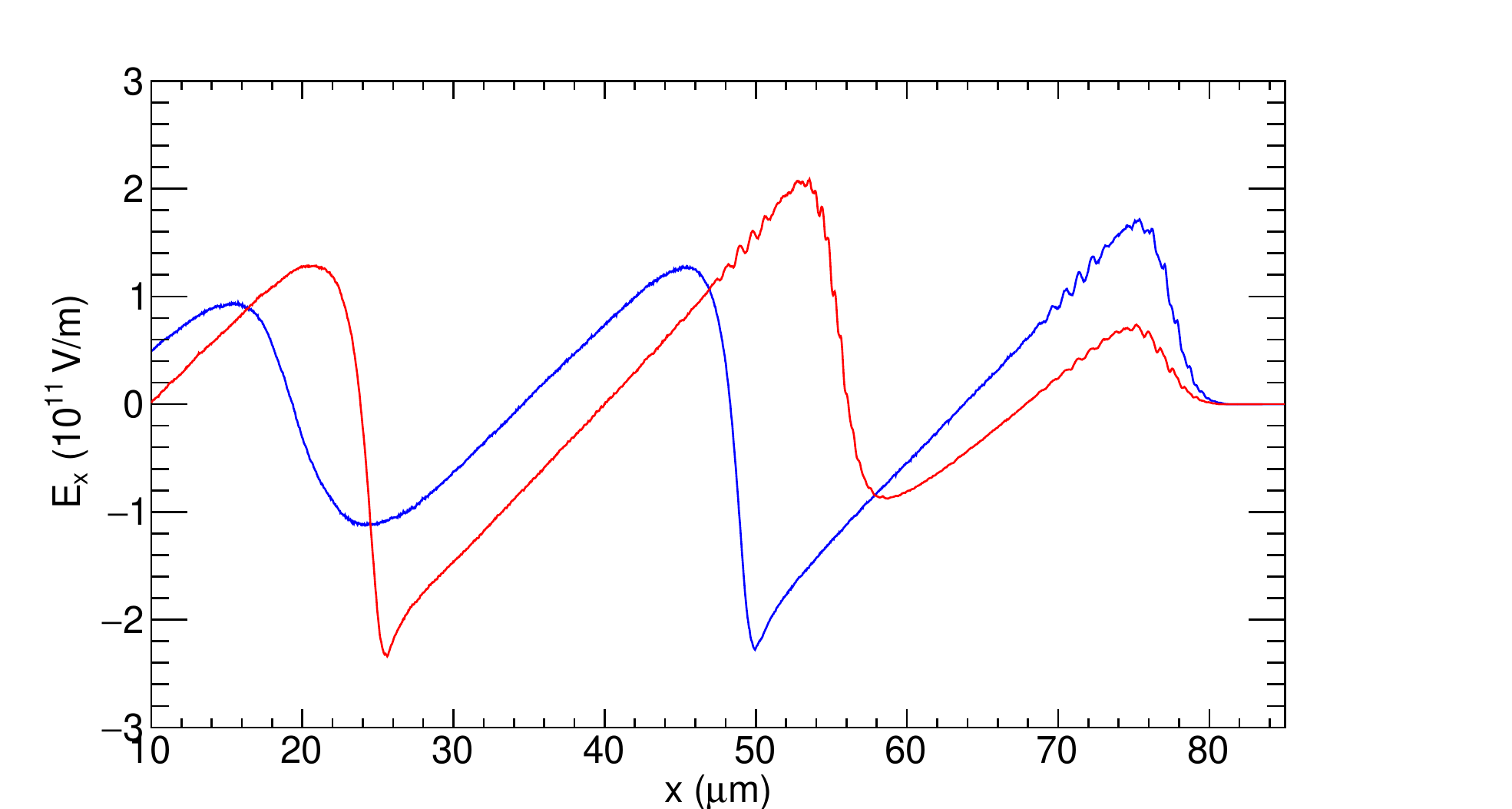}
 	\caption{\it \small Comparison of the longitudinal component of electric field between single-pulse (blue line) and double-pulse schemes (red line), at $t=\SI{1500}{\femto\second}$.}
 	\label{field}
 	\vspace{0cm} 
 \end{wrapfigure}

Given the promising characteristics of the double-pulse scheme, it is worth determining the conditions for which the trapped electron beam properties are optimal. First, the optimal delay between the pulses is expected to correspond to the size of the first cavity, or twice its longitudinal radius $2r_b=3.8k_p^{-1}a_{01}^{1/2}$. This has been confirmed by carrying out a series of simulations with different delays, keeping the energy of the laser pulses equally divided, $\textit{i.e.}\, E_2=E_1$. On the other hand, expelled electrons return back to the rear side of the cavity providing surplus concentrated charge for a second pulse to act on in creating a stronger second cavity. This is confirmed in Fig.~\ref{field}, which shows the longitudinal electric field in both cases, where the advantage of double pulse scheme over single pulse is also apparent by the higher electric field strength extended over a longer distance. Overall, this results in a larger longitudinal cavity size of $2r_b=2.33k_p^{-1}({a_{01}+a_{02}})^{1/2}$.


The optimal energy division was determined by performing a further series of five simulations, varying the energy of each laser pulse between $\frac{1}{6}$ and $\frac{5}{6}$ of a constant total energy ($E_1+E_2=2J$). In all simulations the second laser pulse was placed on the rear side of the cavity created by the first pulse as discussed above. As a result, the general optimum energy fraction is when the first pulse is sufficiently high to meet the usual condition for the cavity formation, $a_0>2$, supplying an optimal quantity of ionized electrons for the second laser pulse. The rest of the laser energy can then be invested in the second laser pulse.



\begin{wrapfigure}{}{70mm}\centering
	\vspace{-1.5cm} 
	\includegraphics[width=70mm]{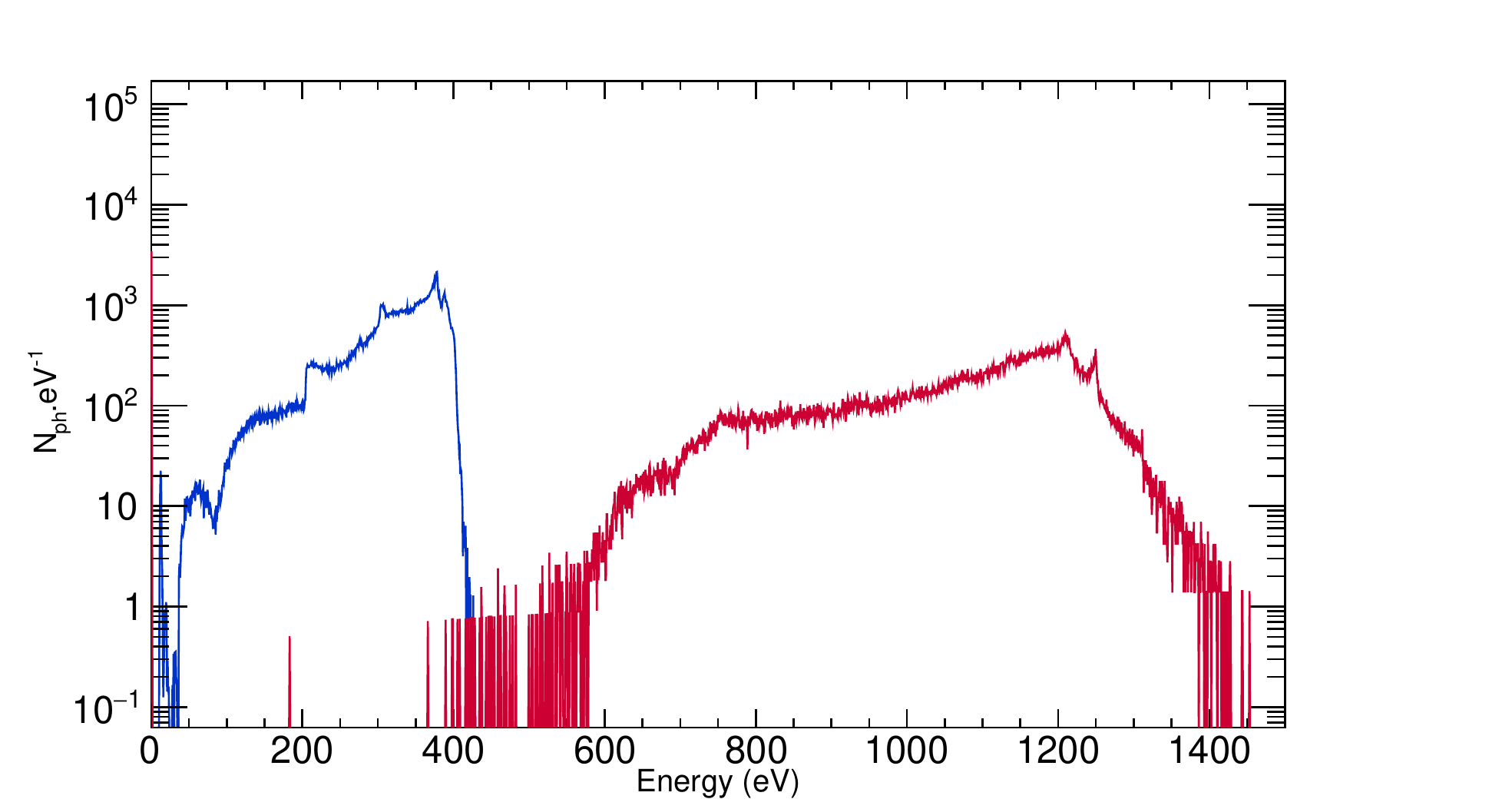}
	\caption{\it \small Comparison of the energy spectra of emitted X-rays from a single betatron oscillation period of accelerated electrons using single (blue line) and double-pulse schemes (red line), for $a_0=3.85$.}
	\label{xcom}
	\vspace{0cm} 
\end{wrapfigure}

Figure~\ref{xcom} makes a quantitative comparison of the energy spectra of emitted betatron X-rays between double- and single-pulse schemes using the formulae quoted previously to estimate the X-ray yield from the trapped electron properties. As mentioned before, for the single-pulse case the trapped electrons in the wake behind the bubble are taken into account. Therefore, higher amount of charge is achieved using single pulse, and consequently higher betatron emission flux, as expected from the radiation equation. However, as it was shown in Fig.~\ref{numdens}-a, the quality of this electron bunch is inferior. Higher electron energy and lower emittance is achieved using the double pulse scheme, contributing to the higher mean X-ray energy.

In summary, we conclude that a tandem-pulse wakefield accelerator in the nonlinear cavity regime offers significant advantages over the conventional single-pulse method, yielding higher electron and betatron emission energies thanks to an enhanced cavity size for the same total laser energy.  We note that it is likely that sequences of 3 or more pulses might permit even greater control over final cavity size and beam energies for the same total pulse energy, extending the original 1D pulse-train concept \cite{FebMulti} to the nonlinear, three-dimensional regime.

This work was carried out in the framework of the \textit{Ju}SPARC project (Juelich Short Pulsed Particle and Radiation Centre) at Forschungszentrum J\"ulich. The authors acknowledge resources provided by the J\"ulich Supercomputer Center (projects JPGI61 and JZAM04). One of the authors (Z.C.) gratefully acknowledges the fruitful discussions with Ilhan Engin and Andr\'e Sobotta as well as visualization support by Jens Henrik G\"obbert.

\end{document}